\documentclass[conference]{IEEEtran}
\IEEEoverridecommandlockouts
\usepackage{graphicx}
\usepackage{mathtools}
\usepackage{graphicx}
\usepackage{comment}
\usepackage{ltablex}
\usepackage{subfig}

\def\BibTeX{{\rm B\kern-.05em{\sc i\kern-.025em b}\kern-.08em
    T\kern-.1667em\lower.7ex\hbox{E}\kern-.125emX}}
\begin{document}

\title{Optimizing travel routes using temporal networks constructed from GPS data}

\author{\IEEEauthorblockN{Tatsuro Mukai, Yuichi Ikeda}
\IEEEauthorblockA{\textit{Graduate School of Advanced Integrated Studies in Human Survivability} \\
\textit{Kyoto University}\\
Kyoto, Japan \\
mukai.tatsuro.l80@kyoto-u.jp,
ikeda.yuichi.2w@kyoto-u.ac.jp}
}

\maketitle

\begin{abstract}
\par Because of the complexity of urban transportation networks and the temporal changes in traffic conditions, it is difficult to assess real-time traffic situations. 
However, the development of information terminals has made it easier to obtain personal mobility information.
In this study, we propose methods for evaluating the mobility of people in a city using global positioning system data.
There are two main methods for evaluating movement.
One is to create a temporal network from real data and check the change in travel time according to time zones or seasons.
Temporal networks are difficult to evaluate because of their time complexity, and in this study, we proposed an evaluation method using the probability density function of travel time.
The other method is to define a time-dependent traveling salesman problem and find an efficient traveling route by finding the shortest path.
By creating a time-dependent traveling salesman problem in an existing city and solving it, a traveler can choose an efficient route by considering traffic conditions at different times of the day.
We used 2 months of data from Kyoto City to conduct a traffic evaluation as a case study.
\end{abstract}


\section{Introduction}
\par One of the major issues confronting urban areas is transportation planning \cite{banister2008sustainable}.
Smooth transportation contributes significantly to residents’ life satisfaction, but transportation planning is becoming more difficult because of the global population growth in urban areas.
In addition, the development of tourism has increased the number of visitors to these areas, and a  known tourism problem has emerged \cite{goodwin2017challenge}.
This negatively affects both visitors and residents \cite{goodwin2017challenge}\cite{milano2018overtourism}\cite{seraphin2018over}.
\par Three reasons make road traffic in urban areas difficult \cite{xue2009traffic}.
First, the urban road networks are typically complicated.
Cities are often densely populated with tunnels and elevated structures that, with numerous roads, form a complex transportation network.
Second, travelers may be heading for different destinations.
This creates a great deal of uncertainty in transportation planning as the number of travelers increases.
Third, the road traffic situation changes over time.
Depending on traffic congestion, travelers may change their transportation and traveling route choices, and travel times for the same traveling route can often vary significantly.
\par However, thanks to the development of information terminals in recent years, it has become easier to obtain mobility data and provide mobility information to individuals.
Information terminals, such as smartphones, are now widely used and carried by most individuals at all times.
Therefore, it has become possible to compose big data from the global positioning system (GPS) measured using these devices and use the data to analyze movement \cite{chen2015design}.
\par To analyze human mobility, network science is useful \cite{barabasi2016network}\cite{lu2013theory}.
Networks whose connection status changes with time are called temporal networks \cite{holme2012temporal}.
They are more difficult to evaluate than static networks because of their time-varying nature, but since they represent real-world problems well, they have been used in various fields such as interpersonal communication \cite{eckmann2004entropy}\cite{iribarren2009impact}\cite{panisson2012dynamics}\cite{vazquez2007impact}, transmission to an unspecified number of people using SNS and the web \cite{adar2005tracking}\cite{kumar2005bursty}\cite{java2007we}\cite{kwak2010twitter}, physical contact \cite{wasserman1994social}\cite{eagle2006reality}\cite{isella2011close}\cite{stehle2011high}, and cytology \cite{przytycka2010toward}\cite{lebre2009inferring}\cite{lebre2010statistical}\cite{rao2007inferring}.
Temporal networks are also preferred to represent movement as a network, to capture changes in traffic conditions over time.
Kujala et al. used temporal networks to evaluate the public transportation system in Helsinki \cite{kujala2018travel}.
They constructed the networks from timetables and open map information and evaluated public transportation by focusing on travel time and the number of transfers.
\par Besides, the shortest path problem can be used to find an efficient travel route.
In recent years, the shortest path problem is as a real-world problem with a time component \cite{dorigo2019ant}.
Heuristic solutions to such a problem, such as the time-dependent traveling salesman problem (TDTSP), have been devised \cite{yongqiang2010improved}\cite{fa2019improved}\cite{mavrovouniotis2017pheromone}\cite{mavrovouniotis2016ant}.
\par In this study, we propose two methods: one is to evaluate human mobility using temporal networks constructed from GPS data, and the other is to search for the shortest path by constructing and solving the TDTSP.
In both methods, location information measured from smartphones is converted into a timetable of location transitions called "transfer connections", from which an optimal set of paths to the destination is obtained to construct mobility networks.
In the evaluation of movement using temporal networks, networks are visualized and compared according to seasons, time zones, and whether the moving person is a resident or a visitor.
In a methodology of finding the shortest path, the time weights of the TDTSP are determined from the set of optimal paths, and the ant colony optimization (ACO) method is used to solve the problem.
In the application of the ACO method, a congestion level is obtained from GPS and used in the calculation.
\par The main contributions of this study include the following:

\begin{itemize}
	\item By applying existing methods for constructing temporal networks and evaluating public transportation, we construct temporal networks from the GPS of smartphones.
	With the GPS, it is possible to take into account other means of transportation other than public transportation, and it is expected that the evaluation will be realistic, including delays in transportation because of congestion.
	With the development of information terminals in recent years, GPS has been commonly used, so it can be easily extended to other cities.
	Temporal networks can describe real-world problems more precisely than can static networks, but evaluating temporal networks is difficult.
	In this study, we proposed a new evaluation method for mobile networks using probability density functions.
	
	\item In this study, we determined the time weights of edges in the TDTSP from GPS data based on the method used to create temporal networks.
	Although existing studies have created the TDTSP from GPS data, the weights were determined originally.
	In addition, to apply ACO the TDTSP, we decide a congestion level from the GPS and calculate the transition probability using this level.
	Selecting a route is difficult if the destination is crowded at certain times of the day.
\end{itemize}

\par The rest of the paper is organized as follows. 
In Section \ref{sec:Methodology}, we introduce the basic methodology.
In Section \ref{sec:Improvements}, we discuss improvements to the basic methodology.
In Section. \ref{sec:meth}, we describe the proposed method.
Following that we discuss the experimental results in Section \ref{sec:results}.
Finally, we conclude our study in Section \ref{sec:con}.

\section{Basic Methodology}
\label{sec:Methodology}
\subsection {Temporal network configuration method}
\par Although networks are used in various fields, several real-world networks change over time.
Networks, those in which the edge connections change with time are called temporal networks \cite{holme2012temporal}.
Temporal networks are often more realistic than static networks and have been used in road traffic analysis.
Kujala et al. \cite{kujala2018travel} constructed a temporal network from public transport timetables to model the movement of people.
They used a multi-criteria profile connection scan algorithm (mcpCSA) to construct the network.
This algorithm is used to compute optimal travels in a dynamic public transportation network, which can represent many events occurring between nodes as a temporal network.

\subsubsection{CSA} 
\par The mcpCSA is an extension of the connection scan algorithm (CSA).
The CSA can calculate the fastest time to reach each stop from a given starting point.
\par Instead of using a graph as Dijkstra's algorithm does, this algorithm uses a single array $C$ of timetable connections sorted by departure time.
Upon receiving the origin $p_s$ and departure time $\tau$ as input, labels $\tau(p)$, which represent the shortest arrival time at each stop, are initialized to infinity.
Then $c \in C$ is scanned to determine whether it is reachable.
If it is reachable and the arrival time of $c$ improves the label $\tau(p_{arr}(c))$ of the destination, update the label.
Repeat until you have scanned all of $C$.

\subsubsection{pCSA}
\par An extension of the CSA for more complex scenarios is the profile CSA (pCSA).
This algorithm returns a set of Pareto-optimal (departure time, arrival time) paths from each stop to a certain destination.
Pareto-optimal here means that the departure time is optimized to be slower and the arrival time is optimized to be faster.
\par This algorithm uses an array $C$ of timetable connections sorted in descending order of departure time.
Whengiven a destination $p_t$ as input, it creates the empty set of Pareto-optimal paths for each stop,
$c \in C$ is scanned in turn, and check if $p_t$ is reachable.
If it is reachable and $(\tau_{dep}(c), \tau\ast)$ is Pareto-optimal compared to the set of paths held by the starting point of $c$, it is added to the set.
Here $\tau_{dep}(c)$ represents the departure time of $c$ and $\tau\ast$ represents the arrival time to $p_t$.
Do this until you have scanned all of $C$.

\subsubsection{mcpCSA}
\par The mcpCSA is an extension of the pCSA.
The mcpCSA extends the Pareto-optimal (departure time, arrival time) treated in the pCSA to the multi-criteria Pareto-optimal like (departure time, arrival time, number of transfers).

\subsection{Time Dependent Traveling Salesman Problem}
\par To reduce traffic congestion, it is important to recommend efficient travel routes to facilitate movement.
To determine the travel route, we formulate and solve the TDTSP.

\subsubsection{Problem Formulation}
\par The TDTSP is a shortest path problem on a time-dependent network represented by the directed graph $G = (V, E, W, T)$.
Here $V$ represents the set of nodes, $E$ represents the set of edges, $W$ represents the set of time weights, and the time interval $T$ is the set of time-varying periods.
\par The weight $w(v_i, v_j, \tau) \in W$ depends on the starting point $v_i \in V$, the ending point $v_j \in V$, and the time $\tau \in T$.
\par $Path(v_1, v_n) = [v_1, v_2 . , v_n]$ represents the transition permutation of nodes from $v_1 \in V$ to $v_n \in V$.
\par When the time starting from $v_1$ is $\tau_0 \in T$, the sum of the time weights of $Path(v_1, v_n)$, $f_{\tau_0}(Path(v_1, v_n))$, is calculated recursively as follows,

\begin{align} 
& \begin{array}{c} f_{\tau_0}(v_1,v_2) = w(v_1,v_2,\tau_0) \end{array},\\
& \begin{array}{c} f_{\tau_0}(v_1,v_i) = w(v_{i-1},v_i,\tau_0 + f_{\tau_0}(v_1,v_{i-1})) \\ + f_{\tau_0}(v_1,v_{i-1}) . \end{array}
\end{align}

\par The TDTSP is a combinatorial optimization problem to find a path such that $f_{\tau_0}(Path(v_1, v_n))$ is minimized.

\subsubsection{ACO}
\par The ACO method \cite{dorigo2006ant} can produce a solution to the TDTSP \cite{yongqiang2010improved}\cite{fa2019improved}\cite{mavrovouniotis2017pheromone}\cite{mavrovouniotis2016ant}.

\par In the ACO method, each edge $e(v_i, v_j) \in E$ has a pheromone $\sigma _{i,j}(\tau)$ for each time $\tau \in T$.
The agent $k$ at node $i \in V$ at time $\tau$ calculates the transition probability $p_{i,j}^k(\tau)$ to a next node according to the pheromone,

\begin{equation}
\label{probability}
p_{i,j}^k(\tau ) = \begin{cases} \frac{{{{\left[ {{\sigma _{i,j}}(\tau)} \right]}^\alpha } \cdot {{\left[ {{\eta _{i,j}}(\tau)} \right]}^\beta }}}{{\sum\nolimits_{s \in \Omega} {{{\left[ {{\sigma _{i,s}}(\tau )} \right]}^\alpha } \cdot {{\left[ {{\eta _{i,s}}(\tau )} \right]}^\beta }} }} & {\text{if}}\;j \in \Omega \\ 0 &  {\text{else }} \end{cases}.
\end{equation}

The next node may be randomly determined using a certain probability, to avoid a local solution.
Here $\Omega$ is the node that agent $k$ has not visited yet, $\eta_{ij}(\tau)$ is the heuristic information, and parameters $\alpha$ and $\beta$ are constants that control the relative importance of pheromone versus heuristic information $\eta_{ij}$.

\par The heuristic information $\eta_{ij}$ is calculated using the following equation,

\begin{equation}
\eta _{i,j}(\tau) = {\frac {Q} {w(v_i,v_j,\tau)}}.
\end{equation}

$Q$ represents a constant and ${w(v_i,v_j,\tau)}$ is the weight of $e(i,j)$ at time $\tau$.

\begin{table}[t]
 \begin{center}
   \caption{Number of daily IDs and logs of GPS data per day in each month of 2019.}
   \begin{tabular}{|c||c|c|} \hline
   Month&The number of ID per day&The number of logs per day \\ \hline
   February &31,456&1,672,421\\ \hline
   April &32,498&1,599,644\\ \hline
   \end{tabular}
  \label{fig:summary}
 \end{center}
\end{table}

\section{Improvements}
\label{sec:Improvements}
\par In this study, we have independently improved the method described in Section \ref{sec:Methodology} to suit our purposes.

\subsection{Improvement of pCSA}
\par In this study, we used the pCSA to construct a temporal network.
It is an algorithm to calculate the optimal traveling route in a dynamic public transportation network.
Since this research uses GPS, we processed the GPS data so that the pCSA could use it.
\par Transfer connections are formed from GPS movements of the same terminal.
GPS data show the location by latitude and longitude, but the granularity is too fine, so we divide the map into meshes and form transfer connections by mesh transitions.

\begin{figure}[t]
\begin{center}
\includegraphics[scale = 0.3]{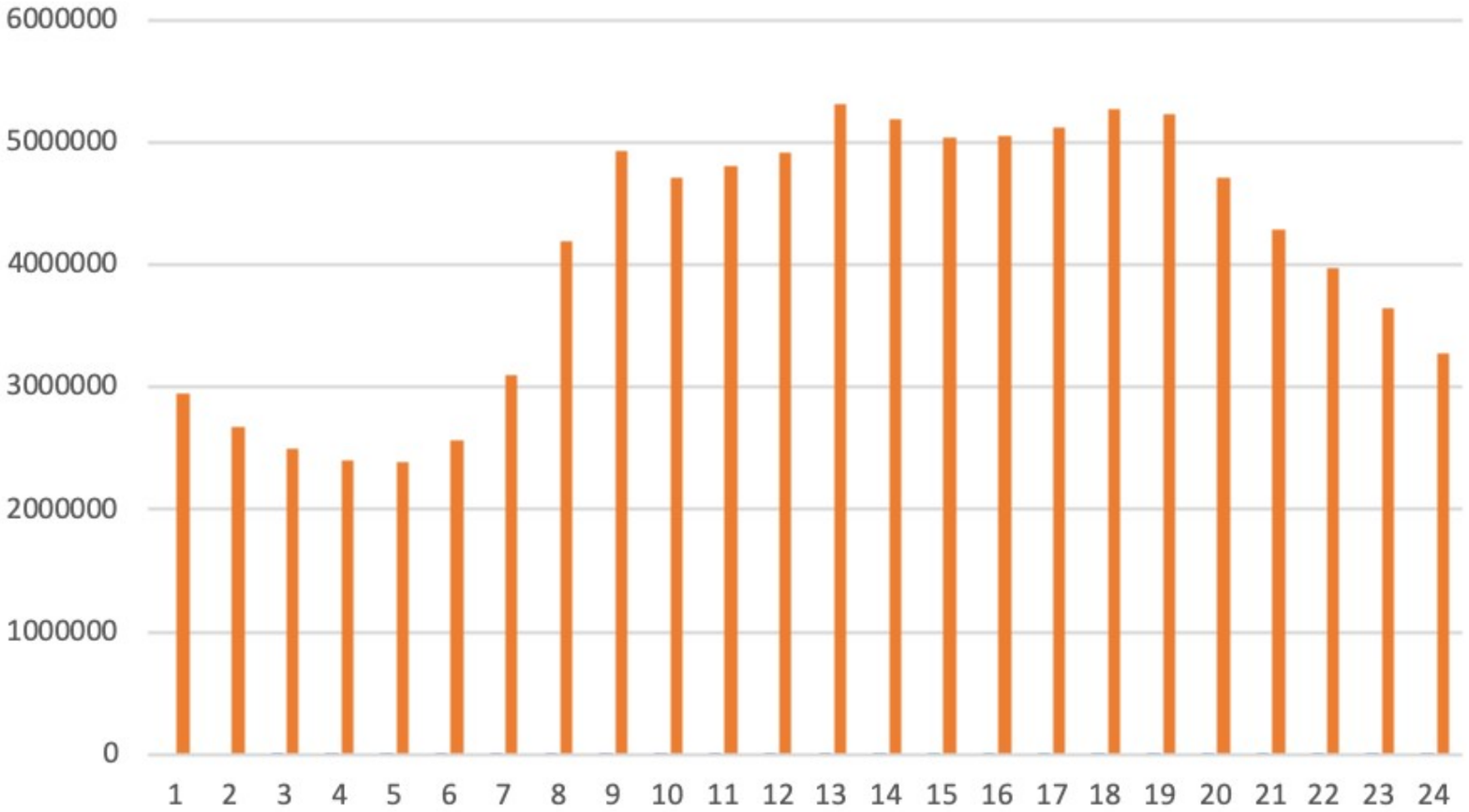}
\caption{The total number of data logs by time zone.
Notably, number of measurements is high during the daytime when people are actively moving and low at night.}
		\label{fig:hourly_log}
	\end{center}
\end{figure}

\begin{figure}[t]
\begin{center}
\includegraphics[scale = 0.3]{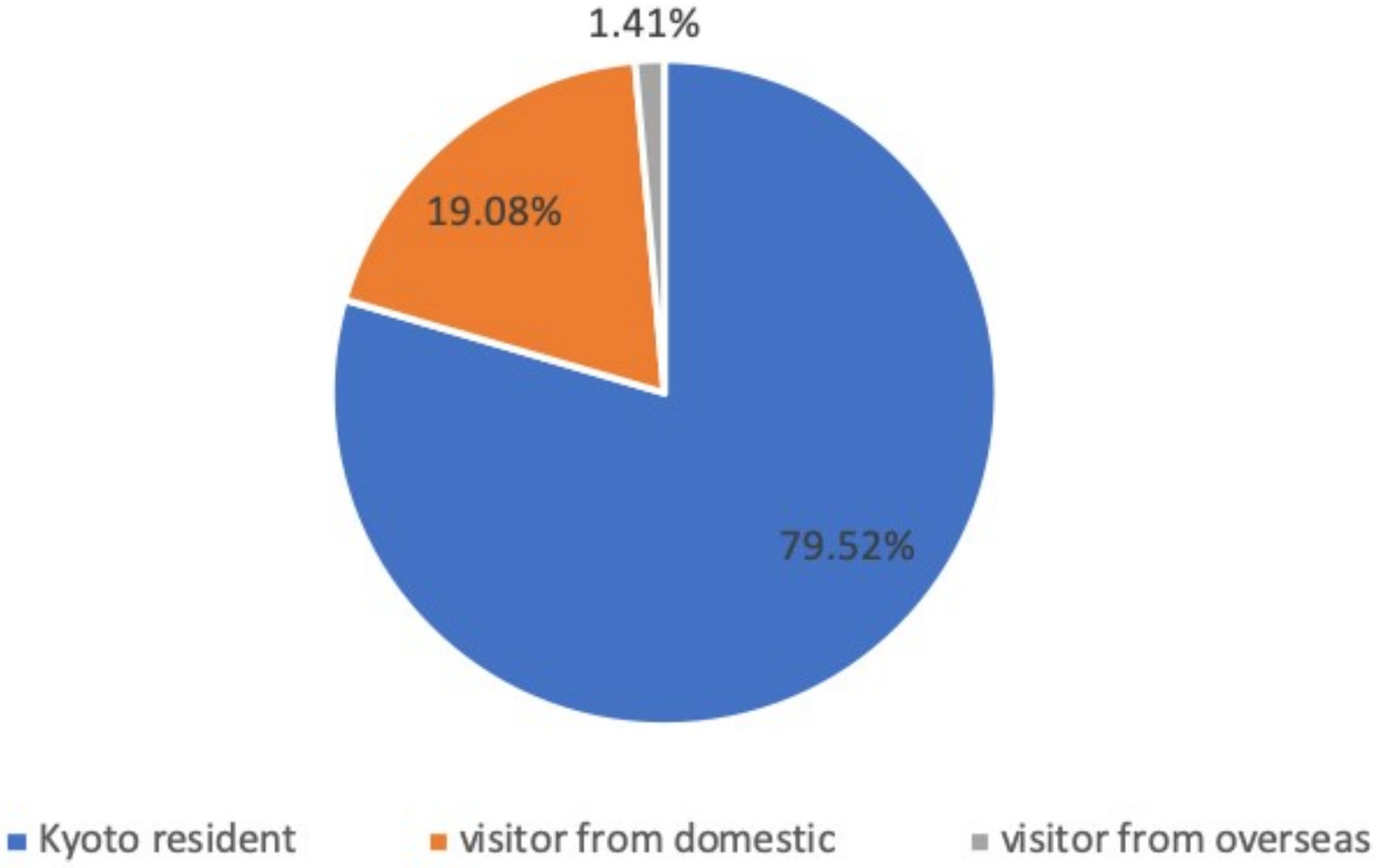}
\caption{Estimated percentage of residential areas in all logs.
Approximately 80\% were Kyoto residents, and 1\% were visitors from abroad.}
		\label{fig:log_attribute}
	\end{center}
\end{figure}

\begin{figure*}[h]
\begin{center}
\includegraphics[scale = 0.6]{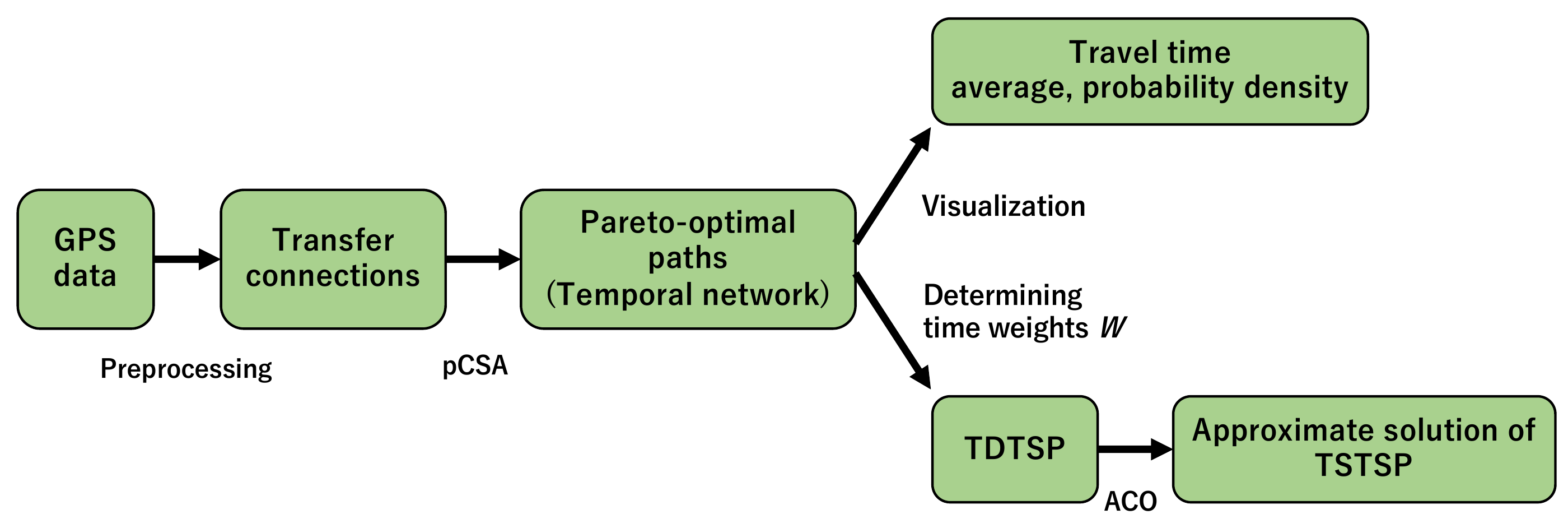}
\caption{Research flow of travel analysis using temporal networks (top) and travel route search using meta-heuristics (bottom).
The flow remains the same until the GPS data are preprocessed and the optimal set of paths is formed using the pCSA.
In the movement analysis using temporal networks, a set of optimal paths is visualized in a form that can be analyzed.
In travel route search using meta-heuristics, a set of optimal paths is used to determine the TDTSP.
}
		\label{fig:reserch_flow}
	\end{center}
\end{figure*}

\subsection{Improvement of ACO}
\ In this study, we set a congestion level and obtain the pheromone for each period from the congestion level.
\par The pheromone $\sigma _{i,j}(\tau)$ is given as follows:

\begin{equation}
{{\sigma _{i,j}}(\tau )} = \frac {\sigma _{0(i,j)}}{\theta_j(\tau)}.
\end{equation}

\par $\sigma_{0(i,j)}$ is the basic pheromone of $e(i,j)$, and $\theta(\tau)$ is the congestion level of the destination at time $\tau$.
\par Congestion is given as a number between 0 and 1, and the more congested a location is, the more difficult it is to be chosen.
Congestion is expressed by the following equation,

\begin{equation}
\label{theta}
 \theta_j(\tau) = {\frac {l_j (\tau)}{l_{j(max)}}}.
\end{equation}

\par We can detect the number of people at node $j$ for each time from the GPS data.
$l_{max}$ is the number of people at node $j$ per time when the number of people staying at node $j$ was the highest, and ${l (\tau)}$ is the number of people at node $j$ during time $\tau$.
\par The update of the basic pheromone is represented by the following equation,

\begin{equation}
{\sigma _{0(i,j)}} = \rho \cdot {\sigma _{0(i,j)}} + \Delta {\sigma _{i,j}}.
\end{equation}

\par $\rho$ is the rate at which the pheromone remains after evaporation.
The increment of pheromone $\Delta {\sigma _{i,j}}$ is determined by ant $k$,

\begin{align}
& \Delta {\sigma _{i,j}}= \sum\nolimits_{k = 1}^m {\Delta \sigma _{i,j}^k}, \\
& \Delta \sigma _{i,j}^k = \begin{cases} \frac{Q}{{\gamma ^k}} & {\text{if}}\;{\text{ant}}\;k\;{\text{pass}}\;{\text{the}}\;{\text{road}}\;e\left({i,j}\right) \\ 0 & {\text{else }} \end{cases}.
\label{final}
 \end{align} 
 
 $Q$ is a constant and $\gamma ^k$is the total moving cost of ant $k$.

\section{Methodology}
\label{sec:meth}
\par With the development of information terminals, it has become possible to track individual movements.
Even handheld ,including smartphones, can use the GPS.
\par In this trend, it is becoming possible to analyze big data of personal mobility data.
It is expected that with large amounts of data, mobility can be evaluated in greater detail, such as when and where people are more likely to move.
\par In this study, we propose methods for analyzing mobility using temporal networks and creating the TDTSP to search for the shortest path with meta-heuristics, using actual GPS data of individuals.

\subsection{GPS data}
\par We used GPS data provided by Agoop Inc.
These data were obtained from users of a smartphone application who gave their consent.
The data are linked to a daily ID that identifies a device and allows it to be tracked for 1 day.

\par This study used 58 days of data in February and April 2019 measured in Kyoto.
In 2019, February was the month when Kyoto had the fewest number of tourists, and April was the month when Kyoto had the highest number of tourists.
Table \ref{fig:summary} shows the amount of data measured in each month.
The number of terminals used for the measurement was higher in April, but the number of logs was lower.
The GPS data used in this study is obtained when the user performs some action when the application is launched. 
While the application is running in the background, it measures when a user moves significantly or stays in facilities, or after a certain amount of time, depending on the OS.
Since the number of logs depends on user activities, it is difficult to determine the cause of the low number of logs in April.

\par Figure \ref{fig:hourly_log} shows the number of logs measured by the hour for all data.
It shows that the number of logs is low during the late-night hours when people are immobile, and the number of logs increases as people become mobile around 8:00.

\par These data are also accompanied by an estimate of the residence location of the terminal owner.
The percentage of all logs classified by immigration is shown in Fig. \ref{fig:log_attribute}.
Twenty percent of visitors were from outside Kyoto Prefecture, and 1\% of all visitors were foreigners.

\begin{figure*}[tb]
\begin{center}
\includegraphics[scale = 0.9]{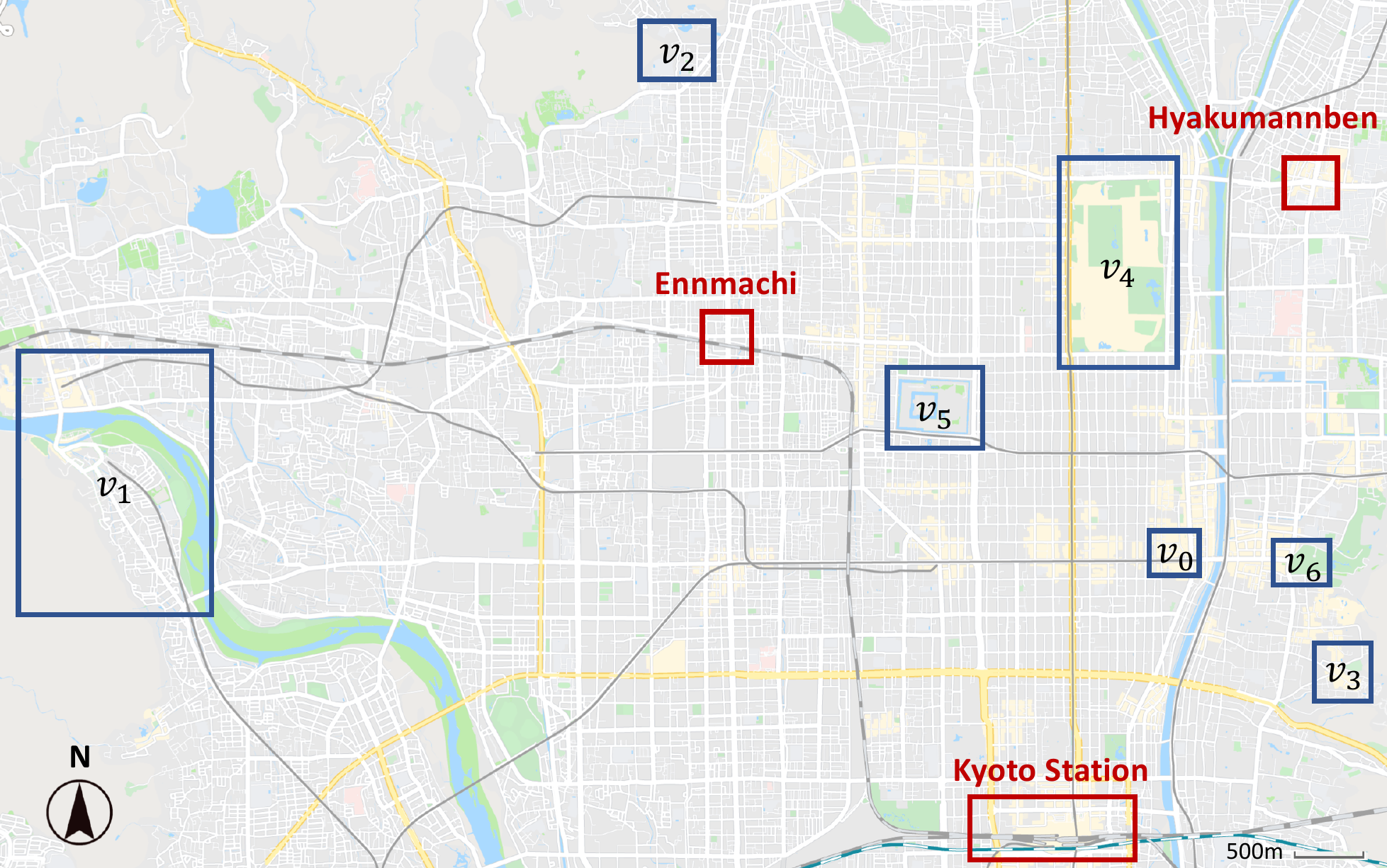}
\caption{Map of Kyoto City showing the locations used in the case study.
The points surrounded by red boxes are the points handled in the temporal network movement analysis, and the points surrounded by blue boxes are the points taken as nodes in the TDTSP.
}
		\label{fig:map}
	\end{center}
\end{figure*}

\subsection{Movement analysis with temporal network}
\par We constructed a temporal network from GPS data and analyzed the movement of people.
Figure \ref{fig:reserch_flow} shows the research flowa, and the details are as follows.

\begin{itemize}
 \item Sort GPS data by the terminal measured.
 \item Transfer connections formed from the same terminal GPS data, and an array C is created by sorting them in descending order of the departure time.
 \item Calculate sets of optimal paths to the destination $p_t$ with the pCSA.
 \item Analyze the resulting set.
\end{itemize}

\par To use the pCSA, GPS data are converted into transfer connections as described above.
In this study, we divided the study area into 50-m square meshes and formed transfer connections between the meshes.
\par From these transfer connections, the pCSA is used to calculate the optimal path to the destination.
The pCSA provides sets of optimized paths to the input $p_t$.
\par For analysis, we compared the travel time by seasons and time zones by creating a probability density function of the travel time to the destination.
This will be discussed in detail in the next section.

\subsection{Travel route search using meta-heuristics}
\par The TDTSP from GPS data was used to calculate the shortest path.
Figure \ref{fig:reserch_flow} shows the research flow, and the details are as follows.

\begin{itemize}
 \item Calculate the optimal path from the GPS data in the same way as described above.
 \item From the obtained sets of optimal paths, determine the travel cost and TDTSP.
 \item Meta-heuristics explore paths with low travel costs around specified points.
\end{itemize}

\par With the pCSA, we find the optimal sets of paths to move the edges in the network.
Afterward, we need to determine the time weight $W$ of the dynamic graph $G$.
\par We now consider how to determine the time weight $w(v_i,v_j,\tau)$ from node $v_i$ to $v_j$.
We use the pCSA with $v_j$ as input, and divide the sets of optimal paths obtained by dividing the sets by time $\tau \in T$.
Considering real-world scenarios, $w(v_i,v_j,\tau)$ should be close to the cost of moving directly from $v_i$ to $v_j$ in that period.
Since there are many paths with long travel times formed by connecting intermediate points using the pCSA, we adopt the shortest travel time in the lower probability 5\% of subgroups as $w(v_i,v_j,\tau)$ in this study.
\par The search for the solution of the TDTSP is calculated on the basis of Eqs. (\ref{probability}) - (\ref{final}).

\section{Results and discussion}
\label{sec:results}
\par Several case studies of movement analysis using temporal networks and shortest travel route search using meta-heuristics are conducted with location information observed in Kyoto City.

\par Figure \ref{fig:map} is a map of Kyoto City showing the locations used in the case study.
The locations surrounded by a red frame represent points used in the temporal network movement analysis.
Movement from Hyakumanben to Kyoto Station and movement from Ennmachi to Kyoto Station are compared and analyzed.
The points surrounded by blue frames represent points taken as nodes for the TDTSP.
We explain the case study in detail.

\subsection{Movement analysis with temporal network}

\subsubsection{Analysis by meantime of movement}
\par We obtained sets of optimal paths from the GPS data using the pCSA.
From the sets of paths, we can obtain the travel time to the destination for each hour.
Here, the travel time includes the waiting time until the next path.
For example, at time $t_0$, if the next path to the destination has a departure time of $t_1$ and an arrival time of $t_2$, the travel time is $t_2 - t_1$ plus the time spent waiting for the path $t_1- t_0$, $t_2 - t_1 + (t_1- t_0) = t_2 - t_0$.

\begin{figure}[h]
\begin{center}
\includegraphics[scale = 0.3]{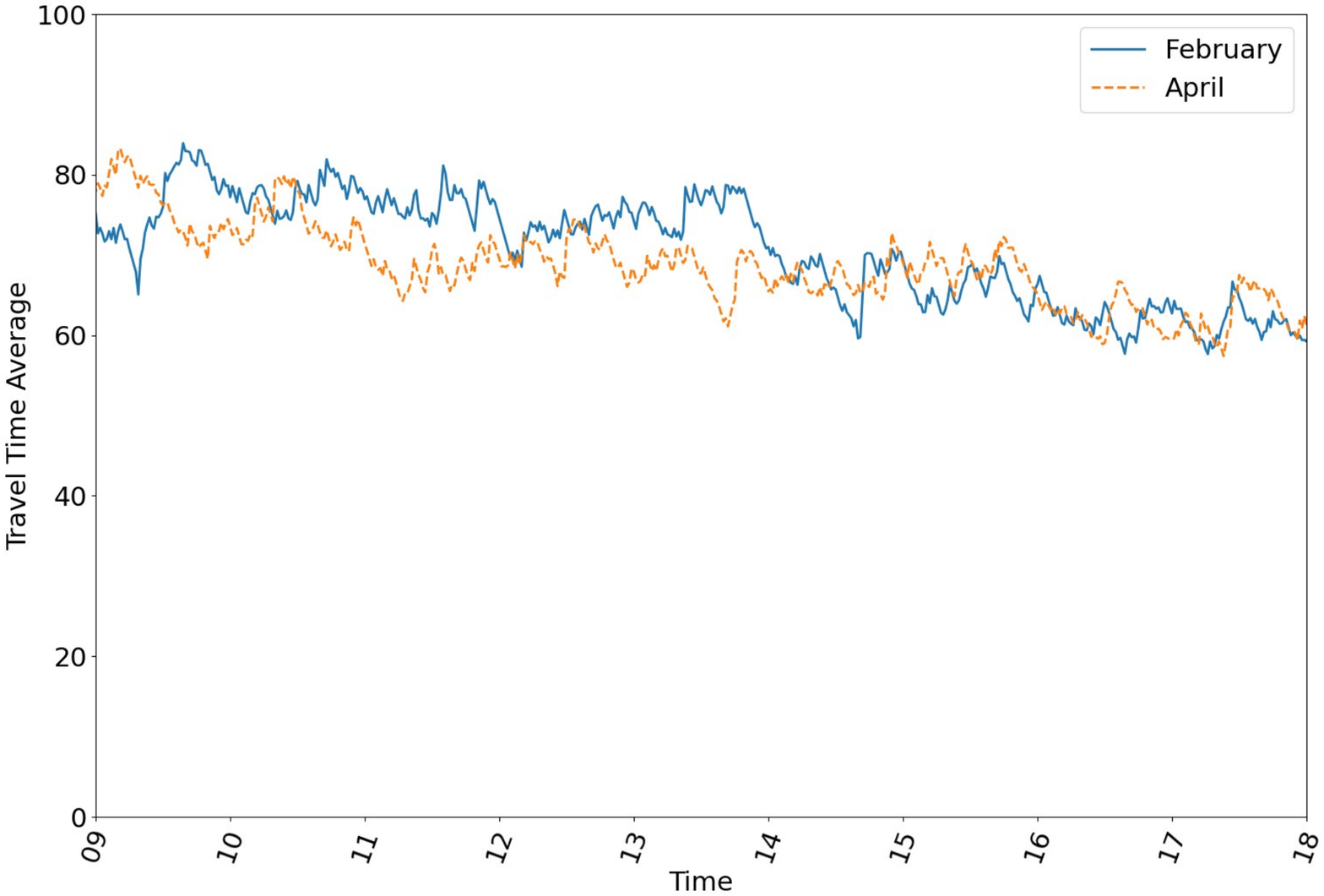}
\caption{The average travel time by time zones from Hyakumanben to Kyoto Station is compared between February (blue line) and April (red broken line).
It shows that the travel time is shorter in April during the morning hours and almost the same from around 14:00 onward.
Note, however, that this includes the waiting time for travel, which may be unrealistic if the amount of data is insufficient.}
		\label{fig:com}
	\end{center}
\end{figure}

\par Figure \ref{fig:com} shows the average travel time from Hyakumanben to Kyoto Station in February and April.
The locations of Hyakumanben and Kyoto Station are shown in Fig. \ref{fig:map}. They are 6.1 km apart by road, so it is expected that it takes approximately 20 min to get there by car.
The average travel time in this study is the monthly average of the hourly travel time for each day.
Since it takes only approximately 30 min to get from Hyakumanben to Kyoto Station by bus, any trip that takes more than 120 min is excluded as an outlier.
The travel time in the morning is shorter in April, and the travel time is almost the same from around 14:00 onwards.
Table \ref{fig:summary} shows that more people are traveling in April, which is counter-intuitive to this result.
However, this result includes the waiting time of the movement.
If only a small amount of movement data is measured, the waiting time will be much longer.
Although the method for evaluating travel time including waiting time is also effective for evaluating public transportation and other transportation with a predetermined timetable, there is a concern that the results will deviate from reality in this research.
Therefore, we will consider an analysis method that is not affected by the waiting time.

\subsubsection{Analysis using the probability density function of travel time}
\par As a case study, we conducted a comparative analysis of the travel time from the Hyakumanben intersection and Ennmachi intersection to Kyoto Station.
The locations of the points are shown in Fig. \ref{fig:map}.
The distance from Ennmachi to Kyoto Station is 5.9 km by road, which is almost the same as the distance from Hyakumanben to Kyoto Station.
However, the trip from Ennmachi to Kyoto Station can be approximately 10 min by train from Ennmachi Station, 150 m away.
\par We can obtain the optimal set of paths to our destination using the pCSA with GPS data.
If we use the pCSA with Kyoto Station as the destination, we can obtain the optimal set of paths from the Hyakumanben intersection and Ennmachi intersection.
A comparative analysis method is used to extract subgroups from the sets that match conditions and compare the density functions of the travel time of the paths included in the subgroups to analyze the trend of absolute travel time without waiting time.

\begin{figure*}[tb]
\begin{center}
\includegraphics[scale = 0.95]{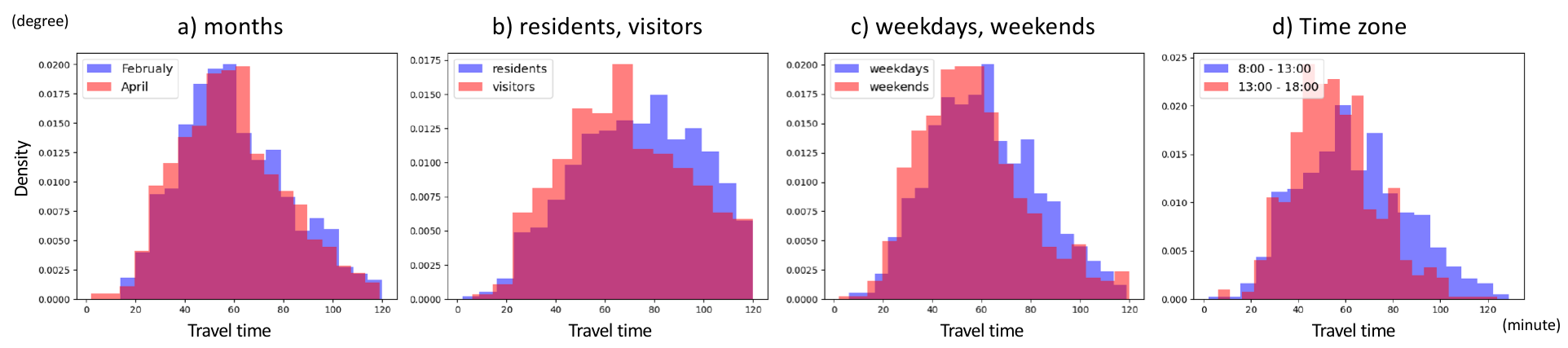}
\caption{A density function is formed for the travel time by classifying the sets of optimal paths for each condition for travel from Hyakumanben to Kyoto Station.
The horizontal axis is the travel time, and the vertical axis is the density.
In (a), the density function is formed for the February and April data.
The density function is formed for Kyoto citizens and visitors in (b), for weekdays and weekends in (c) and for departure time from 8:00 to 13:00 and from 13:00 to 18:00 in (d).
Note that since the data are based on GPS data, the travel time includes not only the effect of traffic congestion but the lag between the start/end of the trip and the time it is measured, as well as the waiting time.
}
		\label{fig:hyaku_histgrams}
	\end{center}
\end{figure*}

\begin{figure*}[tb]
\begin{center}
\includegraphics[scale = 0.95]{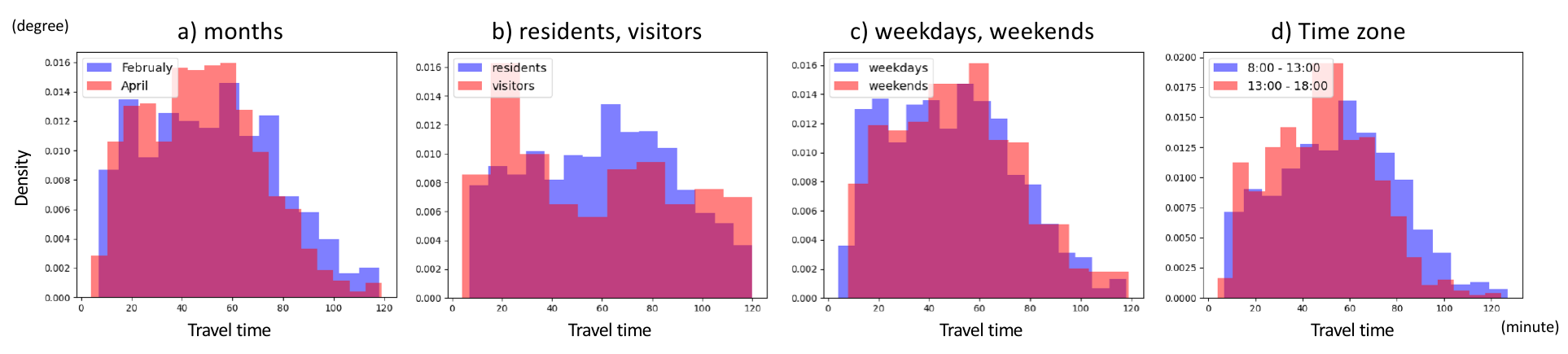}
\caption{A density function is formed for the travel time from Ennmachi to Kyoto Station by classifying sets of optimal paths for each condition.
The conditions in (a)-(d) are the same as in Fig. \ref{fig:hyaku_histgrams}.
By comparing this figure with Fig. \ref{fig:hyaku_histgrams}, the characteristics of the traffic in the section are clarified.
}
		\label{fig:ennmachi_histgrams}
	\end{center}
\end{figure*}

\par Figure \ref{fig:hyaku_histgrams} shows the travel time from Hyakumanben to Kyoto Station, and Fig. \ref{fig:ennmachi_histgrams} shows the travel time from Ennmachi to Kyoto Station.
The density functions are measured for (a) February and April, for (b) Kyoto citizens and visitors, for (c) weekdays and weekends, and for (d) departure times between 8:00 and 13:00 and between 13:00 and 18:00.
\par In 6(a) and 7(a), we compare the probability density functions of the travel time in February and April.
Figure \ref{fig:hyaku_histgrams} shows that the peak of the probability density function is formed at a shorter distance in February.
Compared to that in Fig. \ref{fig:com}, which shows the same comparison in the same interval, the overall travel time was longer in February.
In other words, in Fig. \ref{fig:com}, the waiting time for travel is considered a factor that affects the travel time.
\par In (b), we compare the travel time of Kyoto citizens and visitors.
Comparing the conditions in (b) of Fig. \ref{fig:hyaku_histgrams} and Fig. \ref{fig:ennmachi_histgrams}, there are three major characteristics.

\begin{itemize}
\item Both tend to move quickly because of visitor residence.
\item The movement from Hyakumanben and residents from Ennmachi form a peak at a much longer time than expected for travel by a passenger car or public transportation.
\item The density of visitor movement from Ennmachi peaked at a place with a shorter travel time than others.
\end{itemize}

Focusing on the density function of the movement from Hyakumanben, both the movement of residents and that of visitors form a peak when the time exceeds 60 min.
It takes approximately 20 min to reach Kyoto Station by car from Hyakumanben and approximately 30 min by public transportation, which is considerably longer than the results.
In this study, since the routes are calculated using the pCSA with GPS data, the following factors should be considered in addition to traffic congestion and other events that lengthen the travel time.

\begin{itemize}
\item the lag between the start and end of a movement and the time a smartphone measures location information.
\item waiting time, and
\item impacts of detoured or heavily connected travel formed using the pCSA.
\end{itemize}

Particular attention should be paid to the impact of long-distance or heavily connected movements.
For trips with long distances or from places where few people head directly to the destination, the set of optimal paths includes paths formed by connecting many short paths and detouring paths.
These effects are thought to have formed a peak with a travel time longer than the actualtravel time.
In contrast, the density of travel for visitors from Ennmachi peaks at a short distance, which is thought to be because it is not affected by detouring paths and paths with many connections.
Considering the major difference between travel from Hyakumanben and travel from Ennmachi, there is the fact that visitors can travel from the nearest station in Ennmachi to Kyoto Station on the same train line.
Since most of the visitors traveled by train, we can assume that the peak travel time is approximately 20 min.
According to the timetable, it takes 7-11 min to travel from Ennmachi Station to Kyoto Station, but it is thought that the peak is formed by a discrepancy of a few minutes because of the lag until the location information is measured by a smartphone in transit and the effect of waiting time.
\par Next, in (c), we compared the density function created from the weekday data with the one created from the weekend data.
The travel time from Hyakumanben tended to be shorter on weekends.
When we analyzed travel from Ennmachi to Kyoto Station, we found that a travel time of approximately 20 min, when the peak of visitors seen in (b) was formed, is more frequent on weekdays.
It can be inferred that visitors who use this route are mainly business or commuters rather than tourists.
In addition, (a) shows that there is the same level of travel time of approximately 20 min in February and April, which confirms this assumption.

\begin{table*}[t]
 \begin{center}
   \caption{Time weights $W$}
   \label{fig:weight}
   \scalebox{0.9}{
   \begin{tabular}{|c||c|c|c|c|c|c|c|} \hline
   & \multicolumn{7}{|c|}{Destination}\\ \hline
   depature&$v_0$&$v_1$&$v_2$&$v_3$&$v_4$&$v_5$&$v_6$ \\ \hline
   $v_0$ &[0, 0, 0, 0, 0]&[42, 33, 31, 28, 32]&[54, 52, 50, 45, 49]&[52, 37, 38, 35, 33]&[12, 16, 13, 14, 15]&[23, 22, 25, 24, 21]&[4, 4, 6, 6, 7] \\ \cline{1-8}
   $v_1$ &[42, 36, 32, 30, 28]&[0, 0, 0, 0, 0]&[23, 22, 19, 24, 54]&[73, 27, 60, 38, 59]&[33, 34, 31, 33, 38]&[30, 28, 36, 34, 32]&[55, 45, 45, 36, 34]\\ \cline{1-8}
   $v_2$ &[56, 57, 54, 51, 50]&[26, 23, 19, 19, 56]&[0, 0, 0, 0, 0]&[60, 73, 71, 68, 103]&[21, 22, 18, 18, 20]&[19, 17, 21, 20, 40]&[65, 53, 59, 51, 49] \\ \cline{1-8}
   $v_3$ &[38, 36, 35, 30, 30]&[70, 29, 20, 15, 37]&[83, 82, 68, 67, 67]&[0, 0, 0, 0, 0]&[51, 48, 45, 45, 48]&[59, 51, 45, 39, 42]&[17, 19, 17, 18, 16] \\ \cline{1-8}
   $v_4$ &[16, 17, 17, 15, 16]&[39, 37, 30, 32, 34]&[27, 21, 20, 16, 19]&[44, 55, 47, 51, 51]&[0, 0, 0, 0, 0]&[4, 5, 5, 4, 4]&[15, 16, 16, 15, 17 \\ \cline{1-8}
   $v_5$ &[31, 26, 25, 27, 27]&[27, 29, 26, 30, 28]&[23, 20, 22, 20, 25]&[58, 58, 58, 57, 58]&[5, 5, 4, 5, 5]&[0, 0, 0, 0, 0]&[32, 30, 32, 26, 30] \\ \cline{1-8}
   $v_6$ &[5, 5, 6, 5, 5]&[51, 49, 49, 36, 44]&[61, 49, 58, 50, 60]&[22, 21, 21, 21, 21]&[21, 17, 19, 15, 17]&[29, 25, 26, 30, 26]&[0, 0, 0, 0, 0] \\ \hline
  \end{tabular}
  }
 \end{center}
\end{table*}

\par In (d), we created a figure for the movements that departed between the hours of 8:00 -13:00 and 13:-- -18:00.
From Fig. \ref{fig:hourly_log}, we know that the movement of people becomes active at 8:00.
Both the movements Hyakumanben and from Ennmachi were found to be shorter from 13:00 to 18:00.
When we focus on the movement from Hyakumanben, we find that the weekday movement of residents from (b) and (c) tends to become longer.
In other words, the movement in this section is influenced by the morning commute to work and school, as well as daily life and business activities.
If we focus on travel from Ennmachi, the travel time of approximately  20 min is more common between 13:00 and 18:00.
In other words, it can be inferred that the train travel by visitors seen in (b) is the result of visitors on business using the train on their way home.
\par Thus, travel analysis using temporal networks allowed us to compare travel times under various conditions.
Because of the complexity of road networks and changes in conditions depending on the time zone, it is becoming more difficult to evaluate urban road traffic.
Therefore, it is desirable to conduct traffic evaluations such as the one demonstrated in this study under various conditions and to use a method that clarifies the characteristics of traffic in each city or section.

\subsection{Explore the shortest travel route}
\subsubsection{Creating the TDTSP}
\label{sec:tdtsp}
\par In this case study, we set up the TDTSP to visit seven locations in Kyoto City and derived the shortest path.
\par The seven locations of the nodes are shown in Fig. \ref{fig:map} : starting point $v_0$ is Shijo Kawaramachi, $v_1$ is Arashiyama, $v_2$ is Kinkakuji, $v_3$ is Kiyomizudera, $v_4$ is Kyotogosyo, $v_5$ is Nijyojyo, and $v_6$ is Yasakajinjya.
Starting from $v_0$ at 8:00, weights are updated every 120 min.
\par We used the pCSA with each node as an input to obtain Pareto-optimal sets of paths.
Afterward, subgroups of travel time are created for each time $\tau$, and the shorter 5\% of all transfers was assigned time weight $W$.
This is done so that the sets of Pareto-optimal paths would not be affected by the detours created by the pCSA or by the waiting time included in GPS data.
The weights are listed in Table \ref{fig:weight}.
The array shows, from front to back, the weights for 8:00 - 10:00, 10:00 - 12:00, 12:00 - 14:00, 14:00 - 16:00, and 16:00 - 18:00.

For example, it takes approximately 25 min to get from $v_0$ to $v_1$ by car if there is no traffic congestion along the way.
According to Table \ref{fig:weight}, the smallest weight is 28 min, which is close to the travel time when there is no traffic congestion.
In this case study, we assumed that the travel route between nodes is direct without any detours.
Therefore, if we take the shorter 5\%, we can set time weights close to the assumption.

\par We checked the number of people staying at each point for each time $\tau$ from the GPS and calculated the degree of congestion according to Eq. (\ref{theta}).
The calculation results are shown in Table \ref{fig:congestion}.

\begin{table}[h]
 \begin{center}
   \caption{Congestion level at each time zone $\theta$}
   \begin{tabular}{|c||c|c|c|c|c|} \hline
   & \multicolumn{5}{|c|}{Time zone}\\ \hline
   &8-10 h&10-12 h&12-14 h&14-16 h&16-18 h \\ \hline
   $v_0$ &0.493&0.686&0.782&0.868&1.00 \\ \cline{1-6}
   $v_1$ &0.790&1.00&0.833&0.714&0.526\\ \cline{1-6}
   $v_2$ &0.757&1.00&0.958&0.975&0.483 \\ \cline{1-6}
   $v_3$ &0.406&0.704&0.839&1.00&0.654 \\ \cline{1-6}
   $v_4$ &1.00&0.778&0.854&0.753&0.701 \\ \cline{1-6}
   $v_5$ &1.00&0.818&0.800&0.806&0.725\\ \cline{1-6}
   $v_6$ &0.603&0.819&0.939&1.00&0.906\\ \hline
  \end{tabular}
  \label{fig:congestion}
 \end{center}
\end{table}

\subsubsection{Route search based on time spent}
\par We have set up the TDTSP that travels around Kyoto City, but in reality, people often stay at each node.
Therefore, we calculate the travel time by varying the time spent at each node.
In a static network, the shortest path does not change even if the time spent is taken into account, but in a temporal network, the shortest path may change if the time spent changes.
\par The maximum stay time is as shown in Table \ref{fig:staytime}, and we obtain the shortest route in each case by varying the multiplier of the stay time.

\begin{table}[h]
  \begin{center}
     \caption{Time spent at each location}
     \begin{tabular}{|c||c|} \hline
      Place & St \\ \hline \hline
      $v_1$ &  150 min \\  \hline
      $v_2$ & 50 min \\  \hline
      $v_3$ & 50 min \\  \hline
      $v_4$ & 90 min \\  \hline
      $v_5$ & 50 min \\  \hline
      $v_6$ & 40 min \\  \hline
    \end{tabular}
    \label{fig:staytime}
  \end{center}
\end{table}

\par We run ACO with 100 agents and 200 iterations to find the shortest path; the obtained results are shown in Table \ref{fig:routechange}.
The route changed when the magnification changed from 0.1 to 0.15, and it remained the same from there.
One of the major reasons for this change seems to be that if we take the [$v_0$→$v_4$→$v_5$→$v_2$→$v_1$→$v_3$→$v_6$→$v_0$] route when the spent time is short, we have to use $v_1$→$v_3$ when the travel cost of this is 73 min.

\begin{table}[h]
  \begin{center}
     \caption{Multiplier for time spent at each location and shortest route}
     \begin{tabular}{|c||c|} \hline
      Multiplier for spent time& Route \\ \hline \hline
      0 &  $v_0$→$v_4$→$v_5$→$v_1$→$v_2$→$v_3$→$v_6$→$v_0$ \\  \hline
      0.05 &  $v_0$→$v_4$→$v_5$→$v_1$→$v_2$→$v_3$→$v_6$→$v_0$ \\  \hline
      0.1 &  $v_0$→$v_6$→$v_4$→$v_5$→$v_2$→$v_1$→$v_3$→$v_0$ \\  \hline
      0.15 &  $v_0$→$v_4$→$v_5$→$v_2$→$v_1$→$v_3$→$v_6$→$v_0$ \\  \hline
      0.2 &  $v_0$→$v_4$→$v_5$→$v_2$→$v_1$→$v_3$→$v_6$→$v_0$ \\  \hline
      0.4 &  $v_0$→$v_4$→$v_5$→$v_2$→$v_1$→$v_3$→$v_6$→$v_0$ \\  \hline
      0.6 &  $v_0$→$v_4$→$v_5$→$v_2$→$v_1$→$v_3$→$v_6$→$v_0$ \\  \hline
      0.8 &  $v_0$→$v_4$→$v_5$→$v_2$→$v_1$→$v_3$→$v_6$→$v_0$ \\  \hline
      1.0 &  $v_0$→$v_4$→$v_5$→$v_2$→$v_1$→$v_3$→$v_6$→$v_0$ \\  \hline
    \end{tabular}
    \label{fig:routechange}
  \end{center}
\end{table}

\par In this way, it is possible to obtain sets of Pareto-optimal paths from GPS data, define the TDTSP from the sets, and then use meta-heuristics to search for the shortest path.

\subsubsection{Increasing the time weight of a particular edge}
\par In traffic in urban areas, it is possible that sudden changes in travel time because of events such as accidents, or traffic restrictions may prevent travel along a particular segment.
Therefore, as the next case study, we consider the case where the time weights of specific edges connecting $v_0$ to $v_6$ are extremely increased.
\par Although this is an extreme case, we will analyze how the travel route changes when the time weights $w(v_i, v_j, \tau), w(v_j v_i, \tau) (v_j \in V , \tau \in T)$ of the edges connected to $v_i$ are doubled owing of a large-scale traffic restriction or accident at the point $v_i$.
After setting the time spent in Table \ref{fig:staytime}, we doubled the time weight for each node and performed ACO with 100 agents and 200 iterations to find the shortest path.
The obtained results are shown in Table \ref{fig:each_ver_weigh}.

\begin{table}[h]
 \begin{center}
   \caption{Nodes $v_i$ that double the time weights $w(v_i, v_j, \tau), w(v_j v_i, \tau) (v_j \in V , \tau \in T)$ and the shortest route.}
   \begin{tabular}{|c||c|} \hline
   Node & Route \\ \hline \hline
   $v_0$ & $v_0$→$v_4$→$v_5$→$v_2$→$v_1$→$v_3$→$v_6$→$v_0$ \\ \hline
   $v_1$ & $v_0$→$v_4$→$v_5$→$v_2$→$v_1$→$v_3$→$v_6$→$v_0$ \\ \hline
   $v_2$ & $v_0$→$v_4$→$v_5$→$v_2$→$v_1$→$v_3$→$v_6$→$v_0$\\ \hline
   $v_3$ & $v_0$→$v_6$→$v_3$→$v_1$→$v_2$→$v_5$→$v_4$→$v_0$\\ \hline
   $v_4$ & $v_0$→$v_4$→$v_5$→$v_2$→$v_1$→$v_3$→$v_6$→$v_0$ \\ \hline
   $v_5$ & $v_0$→$v_4$→$v_5$→$v_2$→$v_1$→$v_3$→$v_6$→$v_0$\\ \hline
   $v_6$ & $v_0$→$v_4$→$v_5$→$v_2$→$v_1$→$v_3$→$v_6$→$v_0$\\ \hline
  \end{tabular}
  \label{fig:each_ver_weigh}
 \end{center}
\end{table}

\par Table \ref{fig:each_ver_weigh} shows that, the travel route change only when the time weight $w(v_3, v_j, \tau), w(v_j v_3, \tau) (v_j \in V , \tau \in T)$ of the edge connected to $v_3$ is doubled, otherwise the route remains the same as when some timea were set in Table \ref{fig:routechange}.
This is because the original route moves from $v_1$ to $v_3$, however, the changed route moves from $v_3$ to $v_1$, and as shown Table \ref{fig:weight}, the travel time can be reduced by choosing the right time to travel.
\par In this way, we found that a user can change the TDTSP arbitrarily.
For example, when deciding on a sightseeing route, people may want to avoid visiting a certain zone at a certain time zone because of congestion.
By maximizing the time weights, it would be possible to select a route that avoids congestion.
In addition to the assumption of tourist routes, various restrictions and conditions are assumed in transportation planning and route recommendation.
Since this research uses GPS data, it is expected that the nodes and conditions of the TDTSP can be tailored to a user's preferred form.

\section{Conclusion}
\label{sec:con}
\par In this study, we built a temporal network using GPS data to evaluate the movement of people and to find the shortest route.
We improved an existing method of creating a network for public transportation and created a network using GPS data obtained from smartphones.
To apply the existing method to GPS data, we divided the map into 50-m square meshes and created a timetable of transitions between the meshes.
We can evaluate real events such as travel by nonpublic transportation and delays in transportation using real data.
For the visualization and evaluation of the network, we took the average travel time and created a probability density function of the travel time to compare travel by seasons and time zones.
\par We made an original contribution to the method for setting the TDTSP and calculation in the solution method to search for the shortest path.
The weights of the TDTSP are determined from the sets of Pareto-optimal paths used to create the temporal network.
The congestion measured from the GDP was used to calculate the transition probability in the ACO method used for solving the TDTSP.
With the recent development of information terminals, GPS data can be obtained more easily, so we can change the nodes of transportation networks to be patrolled and extend our method to other cities.
\par The following are possible future extensions.

\begin{itemize}
 \item formulation and post implementation evaluation of transportation plans
 \item application to vehicle routing problems, and
 \item extension to the TDTSP with time frames
\end{itemize}

Because of the complexity of the study city, it is difficult to understand issues in transportation planning, but the temporal network obtained from GPS can clarify the mobility issues.
The method for finding the shortest path problem can be extended to vehicle routing problems.
In addition, we set the congestion level in the calculation of the ACO method, and it can be used to set a time frame such that if the congestion level exceeds a certain height, travel to that point is restricted.
This is because not only is traffic congestion during travel a problem, but so is the concentration of people at a destination.
In addition, there is a growing demand to avoid crowds as a countermeasure to the recent outbreak of COVID-19 infections, and the system (i.e., our methods) can be expected to be expanded to address this.

\bibliography{master}
\bibliographystyle{IEEEtran}

\end{document}